\documentclass[runningheads]{svmult}

\usepackage{makeidx}   
\usepackage{graphicx}  
\usepackage{subeqnar}  
\usepackage{multicol}  
\usepackage{physprbb}  
\makeindex             

\begin{document}
\title*{Chemically inhomogeneous photoionization modelling of the
  planetary nebula SMC~N87}
\toctitle{Chemically inhomogeneous photoionization model of SMC~N87}
%
%
\titlerunning{Chemically inhomogeneous model of SMC~N87}
%
\author{Yiannis G. Tsamis
\and Daniel P{\'e}quignot}
%
\authorrunning{Yiannis G. Tsamis and Daniel P{\'e}quignot}
%
%
\institute{LUTH, Laboratoire l'Univers et ses Th\'eories, associ\'e au CNRS (FRE
2462) et \`a l'Universit\'e Paris 7, Observatoire de Paris-Meudon, F-92195
Meudon C\'edex, France}

\maketitle              


%

We constructed dual abundance photoionization models of
SMC~N87 along the principles laid out by P{\'e}quignot et al. (2002)
in their modelling of the galactic PN NGC\,6153. Recent
spectroscopic studies have necessitated this
approach if one wants to interpret correctly the faint, heavy element
optical recombination lines (ORLs) seen in the spectra of PNe and
H~{\sc ii} regions. The semi-empirical analysis of Tsamis et
al. (2003, 2004) has shown that in N87 the abundances of C and O,
relative to H, derived from ORLs are a factor of 2--3 larger than those
derived from collisionally excited lines (CELs). This discrepancy can
be explained by introducing chemical inhomogeneities in the nebula in
the form of hydrogen-deficient, cold plasma regions.

SMC~N87 is an almost round nebula with no hints of significant
internal structure and a very faint nucleus (cf. \emph{HST} imaging by
Stanghellini et al. 2003). We modelled the PN by combining 4
spherically symmetric computations with the code {\sc nebu},
using \emph{IUE} and optical line fluxes from the NTT 3.5-m (Tsamis et
al. 2003), AAT 3.9-m (Barlow 1987), and \emph{HST} facilities
(Stanghellini et al. 2003).

We consider a scenario in which the central star at some point in its
evolution ejected into the nebula H-deficient material, which originated as
nucleosynthetic products brought up to the post-AGB stellar surface by
the 3rd dredge up. This H-poor (metal-rich) gas was immersed in the more
`normal' abundance inner regions of the PN, condensing into relatively
dense optically thin knots. These, owing to their high metallicity, rapidly
cooled down to temperatures much lower than the ambient gas,
largely via the emission of infrared fine structure transitions.
The model comprises: (1) a H-deficient sector (C1), with
covering factor $\omega/4\pi$=0.70, optical depth $\tau$=0.31
at 1\,Ryd, outer radius 7.14$\times$10$^{16}$\,cm, and a small filling
factor, $f$;
(2) A usual-composition shell (C2), with $\omega/4\pi$=0.87,
$\tau$=5.5, outer radius 3.07$\times$10$^{17}$\,cm, and $f$=1
from which a region coinciding with C1 is removed; (3) The gas
pressure is the same
in C1 and C2 corresponding to conditions of isobaric cooling for C1;
(4) The nebula is radiation bounded overall. Emission
from low ionization species like [O~{\sc i}] and [O~{\sc ii}] led us
to include peripheral clumps in a sector with $\omega/4\pi$ = 0.13,
and `infinite' optical depth; (5) The primary spectrum is a black
body ($L_{BB}$=2.05$\times$10$^{37}$\,erg s$^{-1}$, $T_{BB}$=63.8\,kK) with
a flux cut-off for frequencies above the He$^{+}$ limit
(4\,Ryd) since no He~{\sc ii} lines are observed.

The mean electron densities and temperatures in C1 and C2 are: ($T_{\rm e}$,
$N_{\rm e}$) = (6300\,K, 33\,710\,cm$^{-3}$) and (12\,950\,K,
2900\,cm$^{-3}$), respectively. The mass of the H-poor gas is
0.0027\,M$_\odot$ vs. 0.365\,M$_\odot$ in the normal
component. Therefore, even though C1 has enhanced abundances of oxygen
and carbon relative to C2 (6 and 21 times solar
respectively, by number, relative to H), the
mean elemental abundances in the PN approximate those
returned by the semi-empirical method for $t^{2}=0$ (cf. Table~1). The
model reproduces exactly the C and O ORLs, which are severely
underestimated in chemically homogeneous models.

\begin{table}
\caption{Elemental abundances in SMC~N87 (relative to H)}
\begin{center}
\renewcommand{\arraystretch}{1.4}
\setlength\tabcolsep{5pt}
\begin{tabular}{lccccccc}
\hline\noalign{\smallskip}
      &He    & C    & N      & O    &Ne    &S    &Ar \\
\noalign{\smallskip}
\hline
\noalign{\smallskip}
avg. Model &0.089 & 8.56 & 7.14   & 8.02 & 7.24 &6.00 & 5.39 \\
$t^{2}=0$ &0.097 & 8.58 & 7.04   & 8.03 & 7.03 &*    & 5.32 \\
\hline
\end{tabular}
\end{center}
\label{Tab1a}
\end{table}

SMC~N87 is poor in carbon, nitrogen and argon with respect to non-type I
SMC PNe and has a rather large ionized mass. Only 65\% of the
original C was converted to N during the 1st dredge-up, instead of the
100\% expected SMC conversion efficiency. The surface enhancement of C during the 3rd
dredge-up was only $\Delta$C/H = 3.56$\times$10$^{-4}$, which is atypical for SMC PNe,
resembling rather that of a Galactic PN. For $\Delta$He/$\Delta$C = 10
(Boothroyd \& Sackmann 1988) however, the predicted He/H ratio is
0.087, in agreement with our model value of 0.089.

A potential difficulty with the scenario of a late stellar ejection
as the origin of the knots is that
the C/O ratio in the H-poor gas (=1.8) is smaller than that in the normal
gas (=3.4), when the opposite might be expected. 
We also find that while the He/H
number ratio in the knots cannot be
larger than $\sim$0.14 (otherwise the He~{\sc i} fluxes are overestimated), it
could take values as low as 0.083, indicating that He, along with H, might
also be depleted in the knots. This opens the way for alternative
explanations as to the origins of the metal-rich gas, which could
include a scenario involving a family of photoevaporating planetesimals in
the inner regions of SMC~N87.

%

\end{document}